# MBInception: A new Multi-Block Inception Model for Enhancing Image Processing Efficiency

*Fatemeh Froughirad [1], Reza Bakhoda Eshtivani [2], Hamed Khajavi [3,*] and Amir Rastgoo [3]*

[1] *Department of Mathematics and Appplications, University of Mohaghegh Ardabili, Ardabil, Iran*

[2] *Department of Computer Engineering, K. N. Toosi University of Technology, Tehran, Iran*

[3] *Department of Civil Engineering, K. N. Toosi University of Technology, Tehran, Iran*

**Abstract**

Deep learning models, specifically convolutional neural networks, have transformed the landscape of image classification by autonomously extracting features directly from raw pixel data. This article introduces an innovative image classification model that employs three consecutive inception blocks within a convolutional neural networks framework, providing a comprehensive comparative analysis with well-established architectures such as Visual Geometry Group, Residual Network, and MobileNet. Through the utilization of benchmark datasets, including Canadian Institute for Advanced Researc, Modified National Institute of Standards and Technology database, and Fashion Modified National Institute of Standards and Technology database, we assess the performance of our proposed model in comparison to these benchmarks. The outcomes reveal that our novel model consistently outperforms its counterparts across diverse datasets, underscoring its effectiveness and potential for advancing the current state-of-the-art in image classification. Evaluation metrics further emphasize that the proposed model surpasses the other compared architectures, thereby enhancing the efficiency of image classification on standard datasets.

**Nomenclatures**

| Abbreviation | Description | Abbreviation | Description |
|---|---|---|---|
| CNN | Convolutional Neural Network | $\theta$ | Weight of model |
| ResNet | Residual Network | $\eta$ | Size of step |
| VGG | Visual Geometry Group | $\hat{m}$ and $\hat{v}$ | Bias-corrected estimators |
| DASNet | Deep Attention Selective Network | $m_t$ and $v_t$ | First and second moments |

* *Corresponding Author,*
*E-Mail Address:* hmdkhajavi@email.kntu.ac.ir; hmdkhjv@gmail.com

| | | | |
|---|---|---|---|
| Cifar | Canadian Institute for Advanced Research | IFDL | Improved Fuzzy Deep Learning |
| MNIST | Modified National Institute of Standards and Technology database | NADAM | Nesterov-accelerated Adaptive Moment Estimation |
| GAN | Generative Adversarial Networks | NAG | Nesterov accelerated gradient |
| IEViT | Input Enhanced Vision Transformer | SNES | scalable natural evolution strategies |

## 1. Introduction

In an era dominated by visual information, where images serve as the cornerstone of communication and understanding, the need for robust and efficient image classification systems has become paramount. From identifying objects in everyday scenes to diagnosing diseases from medical scans, the ability to accurately categorize images is fundamental to countless applications across industries. Enter deep learning, a cutting-edge field of artificial intelligence that has revolutionized the landscape of image classification, through utilizing Convolutional Neural Networks (CNNs) at its forefront.

Deep learning, inspired by the intricate workings of the human brain, has enabled machines to mimic the process of learning from data, paving the way for unprecedented advancements in image analysis. At the heart of this technological renaissance lies the CNN, a specialized neural network architecture meticulously crafted to excel at discerning patterns within visual data.

There are various models in the area of image classification with different characteristics and architectural design. In this article, we mainly focus on the image classification models, which are produced based on the CNN architecture. The Convolutional Neural Network (CNN) represents a widely adopted architectural paradigm in the domain of image classification, renowned for its effectiveness in extracting and learning intricate visual features from images. VGG (Visual Geometry Group), ResNet (Residual Network), and MobileNet, are some of the CNN-based image classification models, where each has its own unique characteristics and design.

### 1.1. Previous Studies

Although image classification is recent and comparatively fresh phenomena, it has a vast field of study and has attracted the attention of researchers and engineers. Hence, there are numerous researches with different approaches in this area.

In contrast to traditional convolutional neural networks (CNNs) which are static and operate in a unidirectional manner, Stollenga et al. (Stollenga et al., 2014) proposed a new architecture called

DasNet. the Deep Attention Selective Network (DasNet), implements dynamic feedback mechanisms reminiscent of those found in biological brains. In contrast to conventional CNNs, DasNet possesses the ability to modify its convolutional filter sensitivities while performing classification tasks. This flexibility is made possible through a feedback structure, enabling the network to iteratively adjust its focus and consequently improve classification accuracy. The feedback mechanism undergoes training via direct policy search within a large parameter space, facilitated by SNES (short for scalable natural evolution strategies). Empirical assessments on datasets such as CIFAR-10 and CIFAR-100 reveal DasNet's superior performance compared to existing cutting-edge models.

Zhang et al (Zhang et al., 2016) introduce RoR, a novel architecture designed to improve the optimization capabilities of residual networks. In contrast to conventional methods, RoR concentrates on enhancing the residual mapping within residual mapping, instead of the original residual mapping. By integrating shortcut connections at different hierarchical levels into conventional residual networks, RoR aims to enhance their learning capacity. Importantly, RoR is applicable to various types of residual networks, leading to significant performance improvements. Experimental results demonstrate the effectiveness and adaptability of RoR, achieving outstanding performance across diverse structures similar to residual networks. In particular, RoR-3-WRN58-4+SD models attain groundbreaking performance on CIFAR-10, CIFAR-100, and SVHN datasets, setting new benchmarks, while RoR-3 models also outperform ResNets on the ImageNet dataset. Affonso et al (Affonso et al., 2017) examine wood board quality classification using image analysis techniques. It compares deep learning, particularly Convolutional Neural Networks (CNNs), against a fusion of texture-based feature extraction methodologies and conventional techniques such as decision tree induction, neural networks, nearest neighbors, and support vector machines. Findings suggest that deep learning outperforms traditional methods, particularly in complex scenarios. Empirical results indicate that the proposed texture descriptor method remains highly competitive compared to CNN across all experiments conducted on the image dataset. The research conducted by Wang et al (F. Wang et al., 2017) introduce the "Residual Attention Network," a convolutional neural network that integrates an attention mechanism into cutting-edge feedforward network architecture for seamless end-to-end training. The system comprises Attention Modules, which produce features that are sensitive to attention. Extensive testing on CIFAR-10 and CIFAR-100 datasets validates the efficacy of each module through experimentation. The Residual Attention Network achieves leading object recognition performance on CIFAR-10, CIFAR-100, and ImageNet datasets. Notably, it surpasses ResNet-200 by enhancing top-1 accuracy by 0.6% with fewer layers and fewer forward FLOPs. Additionally, the experiment highlights the network's resilience to noisy labels.

A study done by Mikołajczyk & Grochowski (Mikołajczyk Agnieszka & Grochowski Michal, 2018) addresses a common issue in machine learning: Insufficient training data or an uneven distribution of classes within datasets. It explores data augmentation as a solution, comparing and analyzing various augmentation methods for image classification tasks, ranging from traditional transformations like rotation and cropping to advanced techniques such as Generative Adversarial Networks (GANs). The paper presents an innovative approach to data augmentation centered around image style transfer, which generates new high-quality images by blending the content of one image with the appearance of others. These augmented images can enhance the efficiency of neural network training by providing additional diverse data for pre-training. Yang et al (X. Yang et al., 2018) utilized advanced deep learning methods to address the task of hyperspectral image classification. Unlike typical computer vision tasks that only consider spatial context, their approach utilizes both spatial context and spectral correlation to improve classification accuracy. They introduce four new deep learning models specifically designed for this purpose: 2-D convolutional neural network, 3-D CNN, recurrent 2-D CNN, and recurrent 3-D CNN. Through extensive experiments conducted on six publicly available datasets, this study compares their models with existing state-of-the-art methods. Findings demonstrate the superior performance of the proposed deep learning models, with particular emphasis on the effectiveness of the new models.

Dino et al (Dino et al., 2020) offers a comprehensive overview of recent advancements in Facial Expression Recognition (FER) techniques, examining efficient methods used in FER systems, including feature extraction and classification techniques. It reviews published research from the last five years, providing a summarized and comparative analysis of the algorithms employed in these techniques. Each method aims to achieve high accuracy in recognizing various facial expressions such as happiness, sadness, anger, surprise, fear, and disgust, utilizing images from widely known databases. Wang et al (P. Wang et al., 2021) conducted a comparison and analysis of traditional machine learning (SVM) and deep learning (CNN) image classification algorithms, using SVM and CNN as illustrative examples. Through experimentation with both large-scale (mnist) and small-scale (COREL1000) datasets, it was observed that SVM achieves accuracies of 0.88 on mnist and 0.86 on COREL1000, whereas CNN achieves higher accuracies of 0.98 on mnist and 0.83 on COREL1000. These findings suggest that while traditional machine learning performs better with smaller datasets, deep learning frameworks exhibit higher recognition accuracies with larger datasets. The initial investigation of Okolo et al (Okolo et al., 2022) focuses on assessing the effectiveness of the Vision Transformer (ViT) model in classifying chest X-ray images. Following this, a new model called Input Enhanced Vision Transformer (IEViT) is introduced and evaluated for its ability to enhance performance specifically on chest X-ray images featuring different pathologies. Through experiments conducted on four datasets containing several ailments, the results indicated a consistent superiority of the IEViT model over the ViT model across all datasets and their variations. Thakur & Panse (Thakur & Panse, 2022) unveiled ELSET, an advanced deep learning model crafted for the rapid classification of satellite images based on specific regions in real-time, targeting particular application domains. ELSET utilizes Google Earth Engine to collect vast temporal datasets and employs segmentation model filters tailored to the application to remove outliers. Image sets are divided using an upgraded CNN model that incorporates VGGNet 19, GoogLeNet, and ResNet V2. By combining these algorithms, ELSET achieves precise identification of image-layered regions with a medium level of complexity. Benhari & Hossseini (Benhari & Hossseini, 2022) presented an enhanced Deep Convolutional Neural Network or DCNN, for early detection of cervical cancer using Pap smear images. It tackles the challenge of classifying similar samples within the DCNN's classification layer. To address this, the study proposes an Improved Fuzzy Deep Learning (IFDL) model, combining a Deep Belief Network. The model effectively manages uncertainty among closely related classes. Experimental results on the Herlev cell image dataset underscore the superior performance of the proposed approach in addressing both two-class and seven-class classification tasks.

In a recent study Hua et al (Hua et al., 2024) proposed MSSAL, a novel approach for PolSAR image classification. MSSAL employs a multichannel committee model, LP module, and EL strategy to effectively utilize limited training data. It iteratively fine-tunes a deep neural network using LP and EL, obtaining target pixels. Finally, the trained model predicts labels for all unlabeled data, outperforming other methods on real-world PolSAR datasets with limited labeled samples.

## 1.2. Key Novelty

In this paper, we introduce a fresh deep learning model for image classification, drawing from Convolutional Neural Network structures. Our approach involves merging multiple blocks of the inception model to create a unique framework. We detail our proposed model in the following section and then contrast it with several well-known CNN-based models such as ResNet (Residual Network), VGG (Visual Geometry Group) and MobileNet.

### 1.3. Dataset Description

The proposed model, along with the other mentioned models are comparatively tested and trained using several benchmark datasets. It must be mentioned that 10 percent of the training data are utilized as validation. Datasets such as Cifar-10, Cifar-100, MNIST and fashion MNIST are selected for this purpose.

**Cifar-10** is a widely used dataset in the field of machine learning and computer vision. The CIFAR-10 dataset contains a total of 60,000 color images measuring 32x32 pixels, divided into 10 classes, each with 6,000 images. These classes encompass various objects such as airplanes, automobiles, different animals, ships, and trucks. The dataset is commonly used for training and testing algorithms in image classification tasks. It's considered a benchmark dataset in the field because it's relatively small and manageable while still presenting a real challenge for machine learning models due to the variety of objects and backgrounds in the images. **CIFAR-100** is another dataset provided by the Canadian Institute for Advanced Research (CIFAR). Similar to CIFAR-10, it's widely used in machine learning and computer vision research. However, CIFAR-100 is more challenging as it contains 100 classes instead of just 10. Specifically, The CIFAR-100 dataset is comprised of 60,000 color images, each measuring 32x32 pixels, and categorized into 100 classes, with 600 images assigned to each class. Each class contains 20 subclasses, making it more fine-grained compared to CIFAR-10. The 100 classes cover a wide range of objects and scenes, including animals, vehicles, household items, and natural objects.

Researchers often use CIFAR-100 for evaluating the performance of algorithms in tasks such as object recognition and classification, due to its increased complexity and diversity of classes compared to CIFAR-10 (Barz & Denzler, 2020; Krizhevsky, n.d.).

The MNIST dataset is a frequently utilized standard dataset within the realms of machine learning and computer vision. It represents the Modified National Institute of Standards and Technology database. The dataset contains a large collection of handwritten digits that have been normalized and centered. Specifically, the MNIST dataset consists of 60,000 training images and 10,000 testing images. Each image is grayscale and has a size of 28x28 pixels, making it relatively small compared to other datasets. The digits in the images range from 0 to 9, representing the ten classes. MNIST is often used as a starting point for learning and experimenting with machine learning algorithms, particularly for tasks like digit recognition, classification, and image processing. It serves as a standard benchmark for comparing the performance of different algorithms and techniques in the field. (Kadam et al., 2020; LeCun Y et al., 2010).

**Fashion-MNIST** is a dataset designed as a drop-in replacement for the classic MNIST dataset, offering a more challenging task for machine learning algorithms. It comprises 60,000 training images and 10,000 testing images, each of which is a grayscale 28x28 pixel image of various clothing items such as shirts, trousers, dresses, shoes, and bags. With 10 classes in total, it serves as a benchmark for evaluating the performance of algorithms in tasks like image classification and pattern recognition. Fashion-MNIST is particularly valuable for assessing the robustness and generalization capabilities of machine learning models when dealing with more complex visual data beyond handwritten digits (Xiao et al., 2017).

## 2. Materials and Methodology

The subsequent section offers a concise overview of the examined models and deep learning architectures. Subsequently, a comprehensive exposition of the proposed model and the methodology employed in this study is provided. Also, the preprocessing accomplished on dataset is explained in the following section.

### 2.1. Data Preprocessing

As noted previously, our dataset is consisting of images with different dimensions. Since the input images must have similar dimensions with three canals, we resized the dimensions of the images to 32×32×3, where the 32×32 is the dimension of Image and the 3 is the third canal we added. Some of the image datasets were grayscale and had only one canal. To make them acceptable to our models, we put three layers of image on top of each other so the canals of input data increases from one canal to three canals.

Also, the input images are altered to vectors before entering them to the models.

### 2.2. Convolutional Neural Networks (CNN)

Convolutional Neural Networks (CNNs) are a class of deep neural networks primarily used in the field of computer vision for tasks such as image recognition, classification, segmentation, and more. CNNs are particularly effective for these tasks due to their ability to automatically learn and extract hierarchical patterns and features from input images.

At the core of CNNs are convolutional layers, which apply a series of learnable filters (also known as kernels or feature detectors) to input data. These filters convolve across the input image, computing dot products between the filter weights and the pixel values within a localized region. This process allows the network to identify basic features like edges, corners, and textures. CNN architectures typically consist of multiple layers, including convolutional layers, activation functions (such as ReLU), pooling layers (such as max pooling), and fully connected layers (Li Zewen et al., 2004).

***Convolutional layers*** serve as essential components within Convolutional Neural Networks (CNNs), a prevalent type of deep learning architecture utilized extensively in tasks related to computer vision. These layers play a pivotal role in feature extraction from input data, typically images, through the application of convolution operations.

Within a convolutional layer, the input data undergoes processing utilizing a collection of adjustable filters, also known as kernels or feature detectors. Each filter constitutes a compact weight matrix that is convolved, involving element-wise multiplication and summation, with localized regions of the input data. This convolutional process yields feature maps, which effectively highlight specific patterns or characteristics present within the input information (Albawi Saad et al., 2017; O'Shea & Nash, 2015). ***Activation*** functions are mathematical operations applied to neuron outputs in neural networks, introducing non-linearities essential for learning complex patterns in data. Widely used functions like sigmoid, tanh, ReLU, Leaky ReLU, and softmax play critical roles in network performance. The choice and adjustment of activation functions are pivotal for effective learning and accurate predictions in neural networks. (Sharma et al., 2020). ***Pooling layers*** within CNNs serve as elements employed to reduce the dimensions of feature maps generated by convolutional layers through downsampling, while preserving essential information.

Typical pooling operations comprise max pooling and average pooling, which respectively select the maximum or average value from localized regions of the feature maps. Pooling layers help to make CNNs more computationally efficient, reduce overfitting, and increase translational invariance (Gholamalinezhad & Khosravi, n.d.; Sun et al., 2017).

***Fully connected layers***, also known as dense layers, are a type of layer in neural networks where each neuron is connected to every neuron in the previous layer. In these layers, each neuron receives input from all the neurons in the preceding layer, and its output is computed using a weighted sum of these inputs, followed by an activation function.

Fully connected layers are typically found at the end of neural network architectures and are responsible for learning high-level features and making predictions based on the extracted features. They are commonly used in tasks such as classification and regression. The weights connecting neurons in fully connected layers are learned during the training process through optimization algorithms like gradient descent. These layers play a crucial role in capturing complex relationships in the data and producing the final output of the neural network (Y. Yang et al., 2020).

The CNN-based models which are studied and compared to our proposed model are described briefly in the following section.

### 2.3. Inception

The Inception model refers to a deep learning architecture known as GoogLeNet, which was developed by researchers at Google in 2014 (Szegedy et al., 2014). The Inception model is designed for image classification tasks, specifically targeting the challenge of computational efficiency and accuracy.

One of the notable features of the Inception model is its utilization of what's called "inception modules." These modules are comprised of multiple convolutional layers utilizing a variety of filter sizes, allowing the model to capture features at various scales simultaneously. This architecture enables the network to learn more diverse and rich representations of the input images, leading to improved performance. Inception models have been widely used in various computer vision tasks, including image classification, object detection, and image segmentation.

### 2.4. Residual Network (ResNet) model

ResNet, is a deep learning architecture that was introduced by researchers at Microsoft Research in 2015 (He et al., 2015). ResNet is widely used in computer vision tasks including image classification. it was developed to address the problem of vanishing gradients in very deep neural networks, which can impede training progress and limit performance.

The key innovation of ResNet is the introduction of residual connections, or skip connections, which enable the network to learn residual mappings. These connections allow the information from earlier layers to bypass some layers and be directly fed into deeper layers. By doing so,

ResNet alleviates the vanishing gradient problem and facilitates the training of extremely deep neural networks.

Residual connections in ResNet are typically implemented as identity mappings, where the input to a layer is added to the output of subsequent layers. This facilitates the network in understanding the residual, delineating the distinction between the input and the desired output, without needing to learn the entire mapping from the beginning. ResNet architectures come in different depths (layers) such as 18-layer ResNet, 34-layer ResNet, 50-layer, 101-layer ResNet and 152-layer ResNet.

### 2.5. MobileNet model

MobileNet is a deep learning architecture specifically designed for efficient inference on mobile and embedded devices with limited computational resources. It was introduced by researchers at Google in 2017 (Howard et al., 2017). MobileNet aims to provide a good balance between model size, computational efficiency, and accuracy, making it suitable for applications such as real-time image classification, object detection, and semantic segmentation on mobile devices.

The key innovation of MobileNet is the use of depthwise separable convolutions, which consist of two distinct layers: *depthwise convolution* which conducts independent spatial filtering for each input channel by applying separate convolutional filters, reducing computational cost compared to traditional convolutions that use a single filter across all input channels. After the depthwise convolution, a *pointwise convolution*, also called 1×1 convolution, is employed. This combines the output channels from the depthwise convolution into fewer channels using 1×1 convolution. This process aids in capturing intricate patterns and inter-feature relationships while preserving computational efficiency.

By utilizing depthwise separable convolutions, MobileNet significantly reduces the number of parameters and computations compared to traditional convolutional neural networks, while still maintaining competitive accuracy on various image classification and object detection tasks.

### 2.6. Visual Geometry Group (VGG) model

The VGG model, short for Visual Geometry Group model, is a deep convolutional neural network architecture introduced by researchers at the Visual Geometry Group at the University of Oxford in 2014 (Simonyan & Zisserman, 2014). The VGG network gained prominence for its simplicity and effectiveness in image classification tasks.

The key characteristics of the VGG model are its uniform architecture and deep stack of convolutional layers. Unlike some earlier architectures that used complex modules, VGG consists mainly of 3x3 convolutional layers, followed by max-pooling layers to reduce spatial dimensions. The use of small convolutional filters allows VGG to learn complex features while maintaining a simple and elegant architecture.

The original VGG network architecture includes several variations with different depths, denoted as VGG-11, VGG-13, VGG-16, and VGG-19, depending on the number of layers. For instance, VGG-16 which is studied in our work, comprises 16 weight layers, including 13 convolutional layers and 3 fully connected layers.

Despite being computationally intensive due to its depth, VGG achieved remarkable performance on various image classification benchmarks, such as the ImageNet dataset. Additionally, the simplicity and modularity of the VGG architecture have made it a favorable choice for transfer learning and as a backbone architecture for other computer vision tasks, including image detection and segmentation.

### 2.7. The proposed MBInception model

In this study, a novel deep learning model is presented, which is called MBInception (Multi-Block Inception). The model is designed based on the Inception model. As it mentioned, the proposed model is created through combination of multiple blocks of inception. Each block of inception is made out of several convolutional layers, and consists of two inception modules.

The input data at first, enters to a convolutional layer with 7×7 stride size, where it goes through batch normalization, activation and 2D maxpooling with pooling filter size of 3×3. Afterwards, the input enters to the main block with n number of filters. Inside the aforesaid block, the input data is fed into another block namely the first block. Where it is received by a 1×1 convolutional layer. The dataset, then enters to the first inception module and tasks such as batch normalization, activation and drop out are implemented. Following that, the input enters to the second inception module of the first block, where after batch normalization, the outputs are concatenated with the aforementioned input of the first block, and then goes through activation. Next, the final output of the first block enters to 3×3 convolution in the main block where again tasks like batch normalization and activation are implemented.

The dataset once again enters to the first block and aforesaid process repeats. Consequently, the final output of the main block is introduced to a main block with 2n number of filters and the aforesaid process repeats. Accordingly, the output enters to main blocks with 4n and 8n number of filters, respectively.

Since the convolutional layers are 2D, the outputs are flattened to be altered into 1D. After drop out, the data is directed to the dense layer.

The Figure 1 in the following, demonstrates the proposed model's procedure briefly and clarifies the aforesaid explanation.

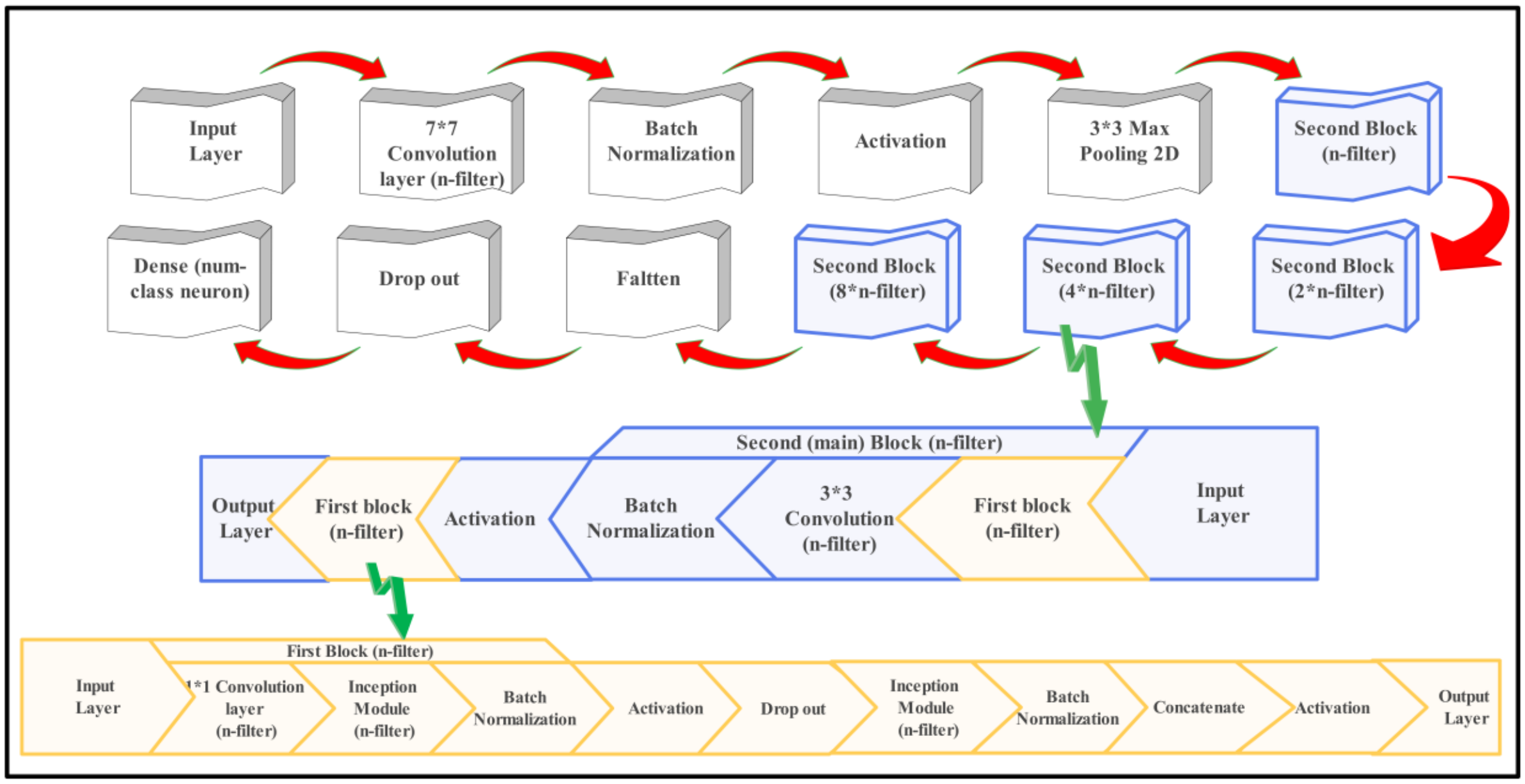

**Figure 1.** the process of the proposed Mult Block inception model

## 2.8. NADAM optimizer

Nesterov-accelerated Adaptive Moment Estimation (NADAM) (Dozat, 2018) is a modified algorithm that combines ADAM optimizer with Nesterov accelerated gradient (NAG). This integrated optimizer accelerates the convergence rate and improves the performance of the models. By incorporating Nesterov momentum into ADAM optimizer, the mathematical expression of NADAM update rule will be as follows (Halgamuge et al., 2020):

$$\theta_{t+1} = \theta_t - \frac{\eta}{\sqrt{\hat{v}}+\epsilon}\left(\hat{m}\beta_1 + G_t\frac{(1-\beta_1)}{1-\beta_1^t}\right) \quad (1)$$

Where $\theta$ is the weight of model. $\eta$ denotes the size of each step and $\epsilon$ is a constant for smoothing, equal to $10^{-6}$. $\hat{m}$ and $\hat{v}$ are two parameters called bias-corrected estimators, where $\hat{m}$ is for first moment and $\hat{v}$ is for second moment, and can be calculated as follows:

$$\hat{m} = \frac{m_t}{1-\beta_1^t} \quad (2)$$

$$\hat{v} = \frac{v_t}{1-\beta_2^t} \tag{3}$$

Where $m_t$ is the estimation of first moment and $v_t$ is the estimation of second moment, also called decaying averages. These values can be obtained using the equations below (Halgamuge et al., 2020):

$$m_t = \beta_1 m_{t-1} + g_t\,(1 - \beta_1) \tag{4}$$

$$v_t = \beta_2 v_{t-1} + g_t^2 (1 - \beta_2) \tag{5}$$

### 2.9. Evaluation Criterions

In order to validate the obtained results from four studied models along with our proposed model, several evaluation metrics are contrived. Criterions such as accuracy, precision, recall and F-1 score are recruited for various purposes to ensure that the acquired results from each model are reliable.

***Accuracy*** quantifies the proximity of obtained results to the expected outcomes, reflecting the degree of correct predictions made by a model. However, accuracy may not be a dependable metric when dealing with unbalanced datasets. Mathematically, accuracy is computed by summing up both true positive and true negative outcomes and dividing them by the total number of positive and negative predictions.

$$accuracy = \frac{true\ positive + true\ negative}{total\ answers} \tag{6}$$

***Precision*** measures the consistency and similarity of results. In essence, high precision indicates that the results are closely aligned with each other. Therefore, precision must exhibit consistency over time. Mathematically, precision is determined by dividing the number of correctly predicted positive outcomes by the total predicted positives, regardless of whether they are true or false.

$$precision = \frac{true\ positive}{true\ positive + false\ positive}$$

***Recall*** assesses the model's capability to accurately predict true positive outcomes. Specifically, it signifies the proportion of true positive answers correctly predicted out of all true positive instances. The mathematical expression for recall is depicted as follows: (Hassan et al., 2022):

$$recall = \frac{true\ positive}{true\ positive + false\ negative} \tag{8}$$

The ***F1-score*** serves as the harmonic mean of precision and recall, particularly valuable when dealing with imbalanced datasets. A high F1 score signifies a well-balanced model performance, where both precision and recall achieve elevated values.

$$f1\ score = 2 \times \frac{precision \times recall}{precision + recall} \tag{9}$$

## 3. Results & Discasion

In this section a comparative approach is presented. Where our novel MBInception model, along with three models including ResNet (Residual Network), VGG (Visual Geometry Group) and MobileNet are tested using several benchmark datasets. All the studied models are created using architecture of inception.

The performance of models on cifar10 demonstrates the following results. Fig. 2 shows the detection results of the selected models along with the proposed model.

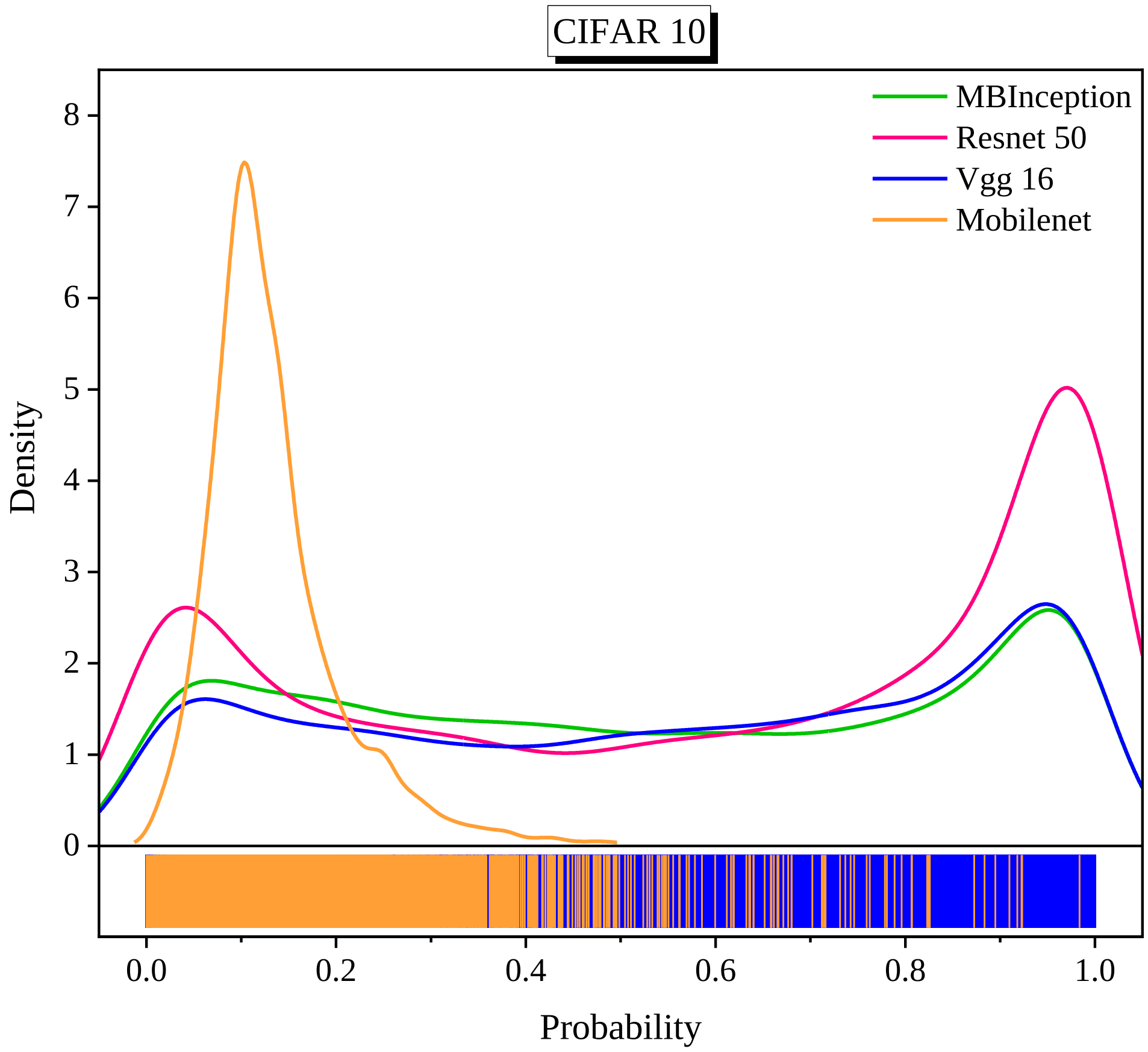

**Figure 2.** Detection results of models on cifar10 dataset

The prediction result of Mobilenet model shows that only fractional number of labels are predicted correctly as 1. The graph shows that the model's performance is so disappointing with the cifar10 dataset. The graph in Fig. 2 shows if the model has predicted the labels thoroughly. There are two classes in datasets namely 0 and 1. If the density of probability is close to 1, it indicates that the model is well-performed and the model's confidence is high. According to Fig. 2, vgg16 was not able to predict all the labels very well. Since the density of the probabilities are so dispersed.
According to Fig. 2, the ResNet model, has predicted more classes close to 1. the graph illustrates that the density of probabilities is mostly accumulated around 1.0. So, the Resnet model outperforms the vgg16 on cifar10.
The prediction result of the proposed model shows that the MBInception model, offers better results than Mobilenet model. Although, compared to other models like vgg16 and Resnet, the proposed model was able to detect less labels close to one. However, since the ResNet predicted more labels as 0, the overall accuracy of the model is lower than the proposed model.
The performance of models on cifar100 which is another benchmark dataset is demonstrated in the following graph.

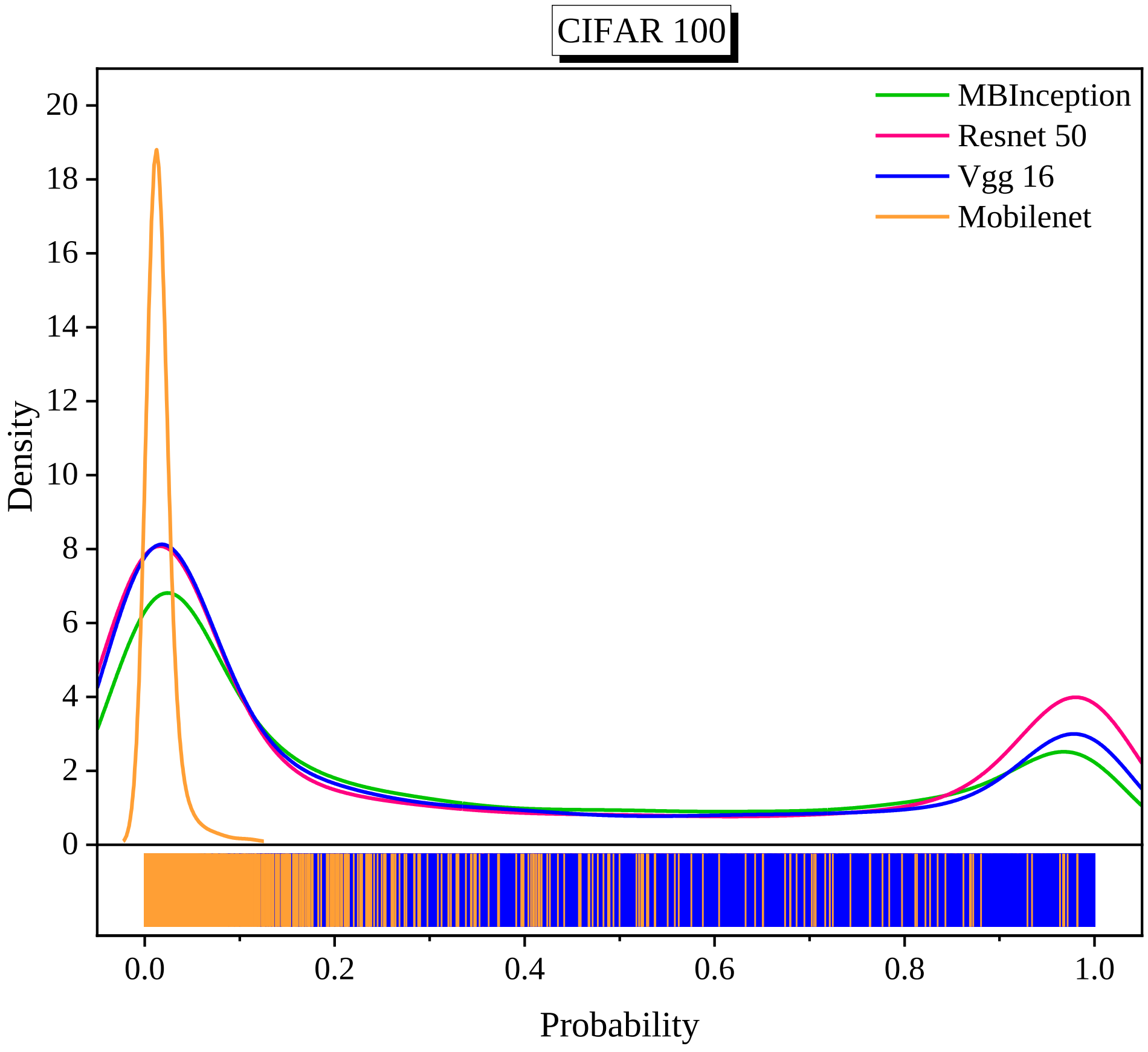

**Figure 3.** Detection results of models on cifar100 dataset

The performance of vgg16 on cifar100 indicates that the model is not able to detect most of the labels precisely. As it can be seen in the Fig. 3, the probability of 0.0 has more density, meaning the model was unable the detect most of the labels as 1.0.

The performance of ResNet on cifar100 depicts comparatively better results than vgg16. Although, the model predicted most of the labels as 0.0; the number of correct detections is higher than vgg16. The graph shows that the Mobilenet is unable to detect the labels correctly and performs infirmly. The density of probability is accumulated in 0.0, meaning low detection results on cifar100 dataset. According to fig. 3, the performance of the proposed MBInception model on detecting labels is similar to vgg16. While, the ResNet shows more promising results on recognizing the labels as 1, but the proposed model has less detected labels as 0. So, the confidence of the models including vgg16, ResNet and MBInception does not have noticeable difference.

The performance results of the models on MNIST dataset, are presented in the following.

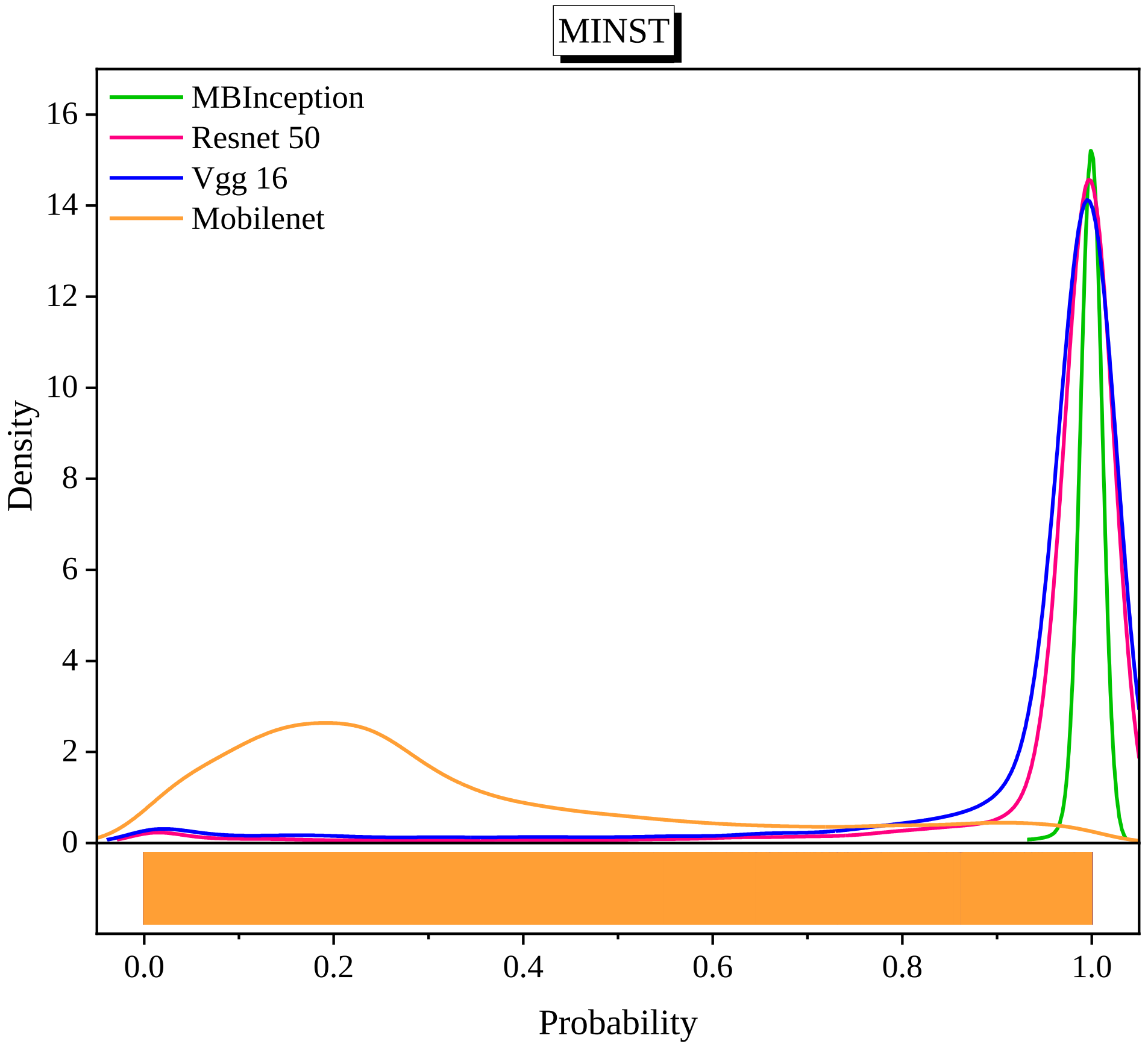

**Figure 4.** Detection results of the models on mnist dataset

According to Fig. 4, the results of vgg16 on MNIST dataset, shows acceptable results. The model prevailed in detecting most of the labels close to 1.0. Density of probability indicates the model performance has high confidence, meaning the model succeeded in detecting the labels exactly.
Similar to vgg16, the ResNet model also detected most of the labels correctly. Also, as it can be seen in Fig. 4, number of correctly detected labels by ResNet is more than vgg16. Probability of ResNet is denser around the 1.0, indicating that the ResNet is more confident model.
The results of the proposed model presented in Fig. 4 demonstrates that, the new model is well-performed in detecting labels. According the graph, the model prospered to detect most of the models correctly and more than both vgg16 and ResNet. Besides, the density indicates that the model was able to recognize most of the labels exactly as 1.0. In other words, the model confidence is high.
The detection result of Mobilenet shows unsatisfying performance of the model. Same for other datasets, the model was not able to detect most of the labels and probability densities are so dispersed with the lowest accuracy among the models.
The following graph depicts the results of detections accomplished by models on fashion-mnist datasets.

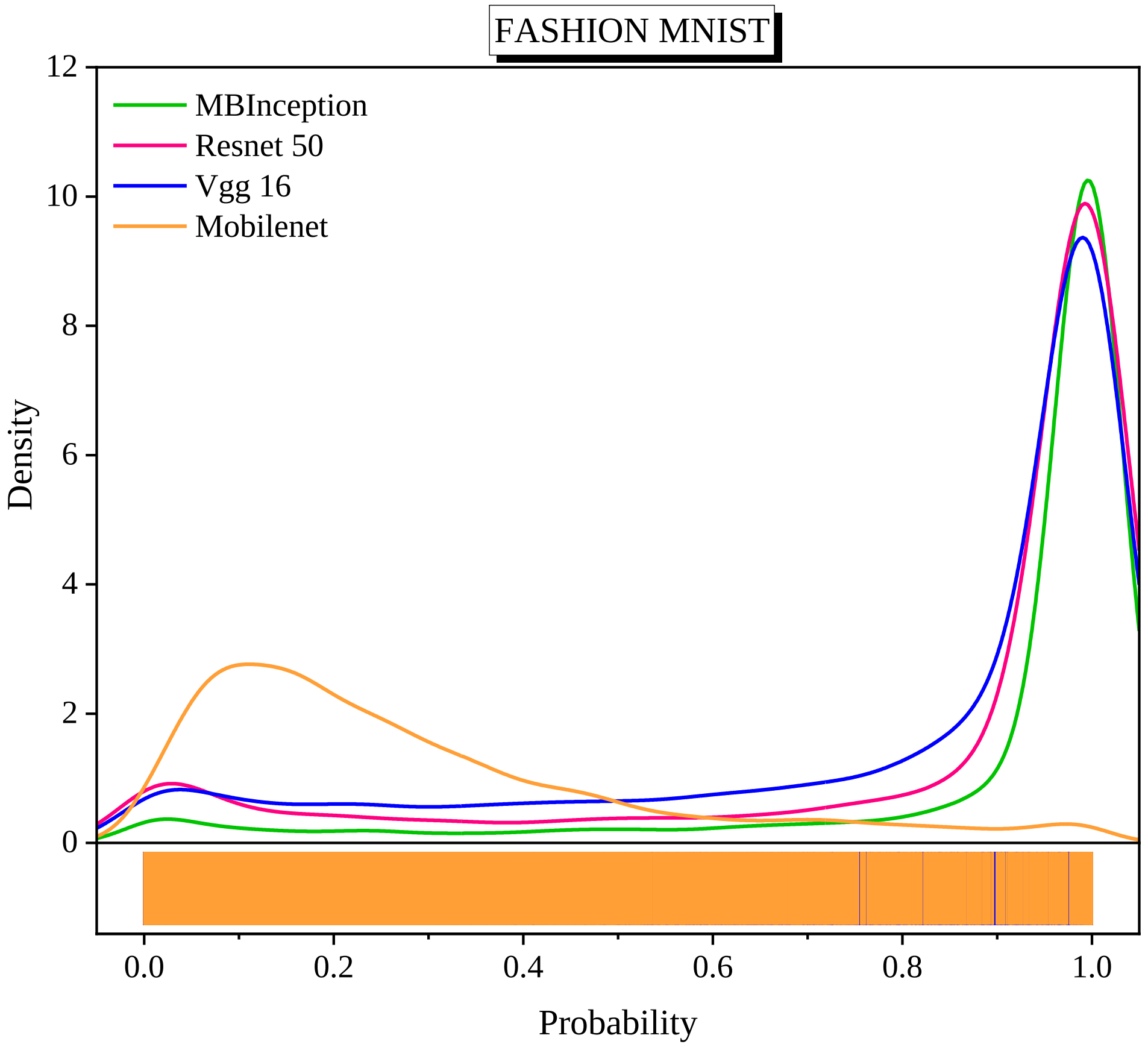

**Figure 3.** Detection results of the models on fashion-mnist dataset

The Fig. 5 shows that the vgg16 model detected most of the labels in fashion-mnist correctly. Although, there are a few labels that are not detected by model. Besides, Fig. 5 demonstrates that the ResNet model is succeeded on detecting most of the labels as the density of probabilities are mostly close to 1.0. By comparing ResNet to vgg16, it is distinct that the ResNet model detected more labels, indicating the model is more confident.

The result of our new model presents promising outcome. According to comparative graphs, the model overcomes other models in detecting the labels, noticeably with more labels as 1 and less labels as 0. The graph denotes that the MBInception model was able to detect most of the labels exactly with less slips, compared to the other models. Furthermore, the density indicates the reliability of model, hence the model detected the exact value of labels.

The Fig. 5 shows that the Mobilenet model detects small number of labels, denoting its unacceptable performance. So that, the model detected most of the labels as 0 while, only small number of labels are detected correctly.

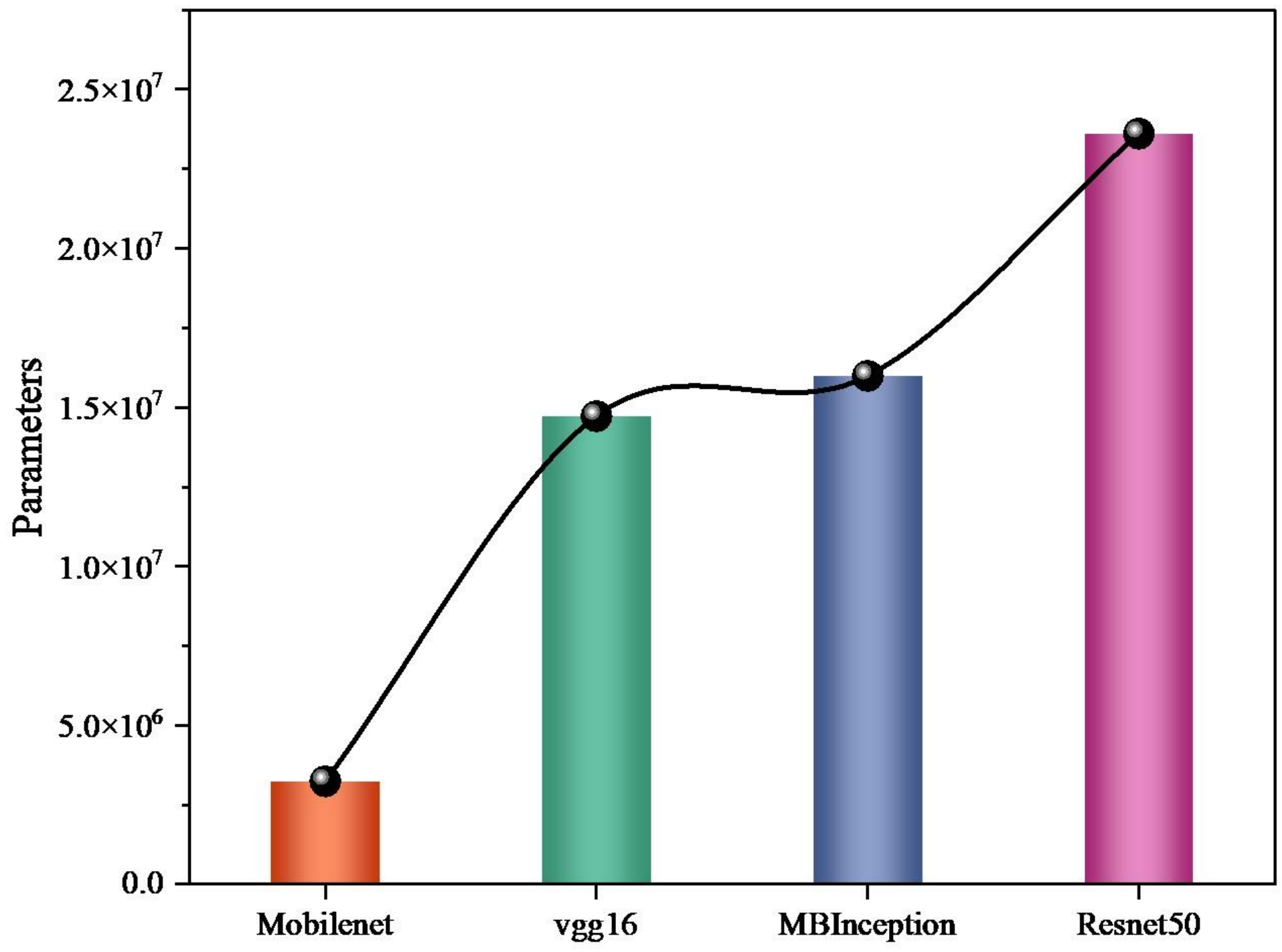

**Figure 6.** Number of parameters of the proposed model and the studied models

As it can be seen in Fig. 6, our proposed model has more parameters than vgg16 and mobilenet, but has less parameters than ResNet model. This indicated that, although the proposed model doesn't have the most parameters among the models, it outperforms the models in most of the tasks performed on different datasets.

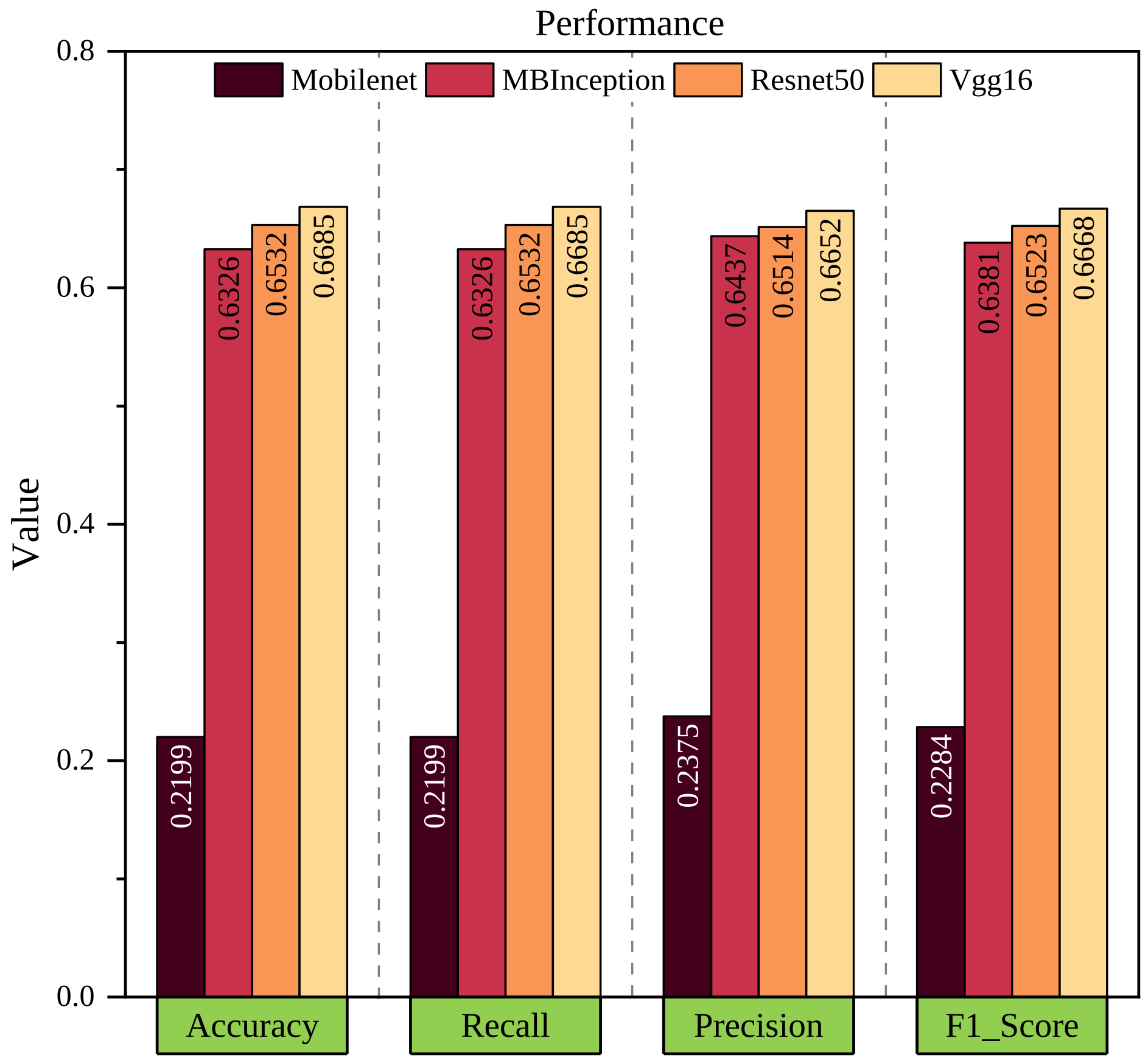

**Figure 7.** the results of evaluation metrics of models for cifar10 dataset

The results of evaluation criterions such as accuracy, precision, recall and f1-score showed in Fig. 7 indicates that vgg16 outperforms other models with promising results. ResNet and the novel proposed model are second and third in performance based on cifar10 dataset. Also, Mobilent shows remarkably low performance compared to the other models.

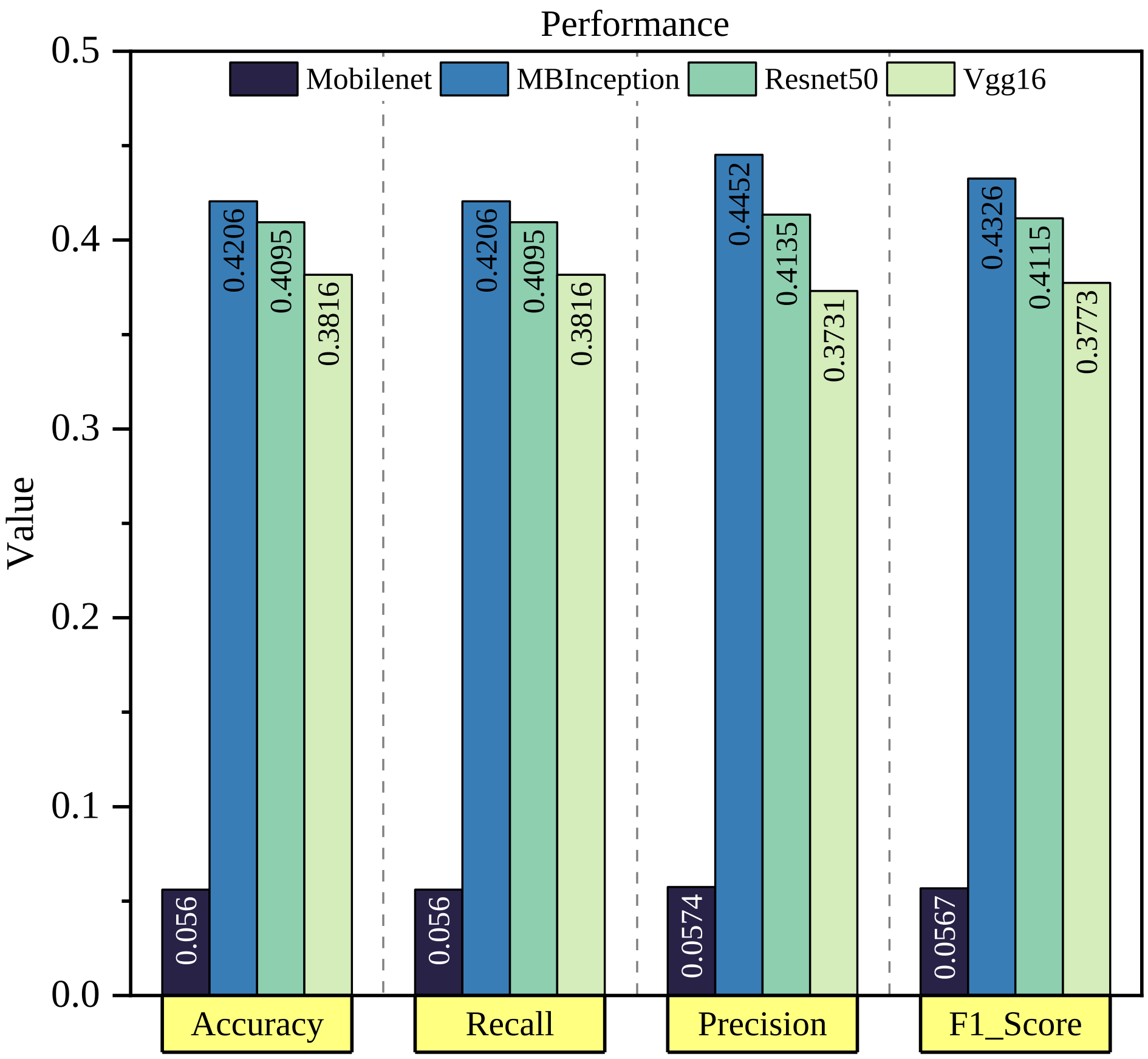

**Figure 8.** the results of evaluation metrics of models for cifar100 dataset

As it can be seen in Fig. 8, our proposed model shows superior results according to the metrics on cifar100, especially in f1-score and precision. Besides, ResNet and vgg16 are second and third, respectively. The results of Fig. 7 and Fig. 8 indicate that the proposed MBInception model performs better in more complicated datasets.

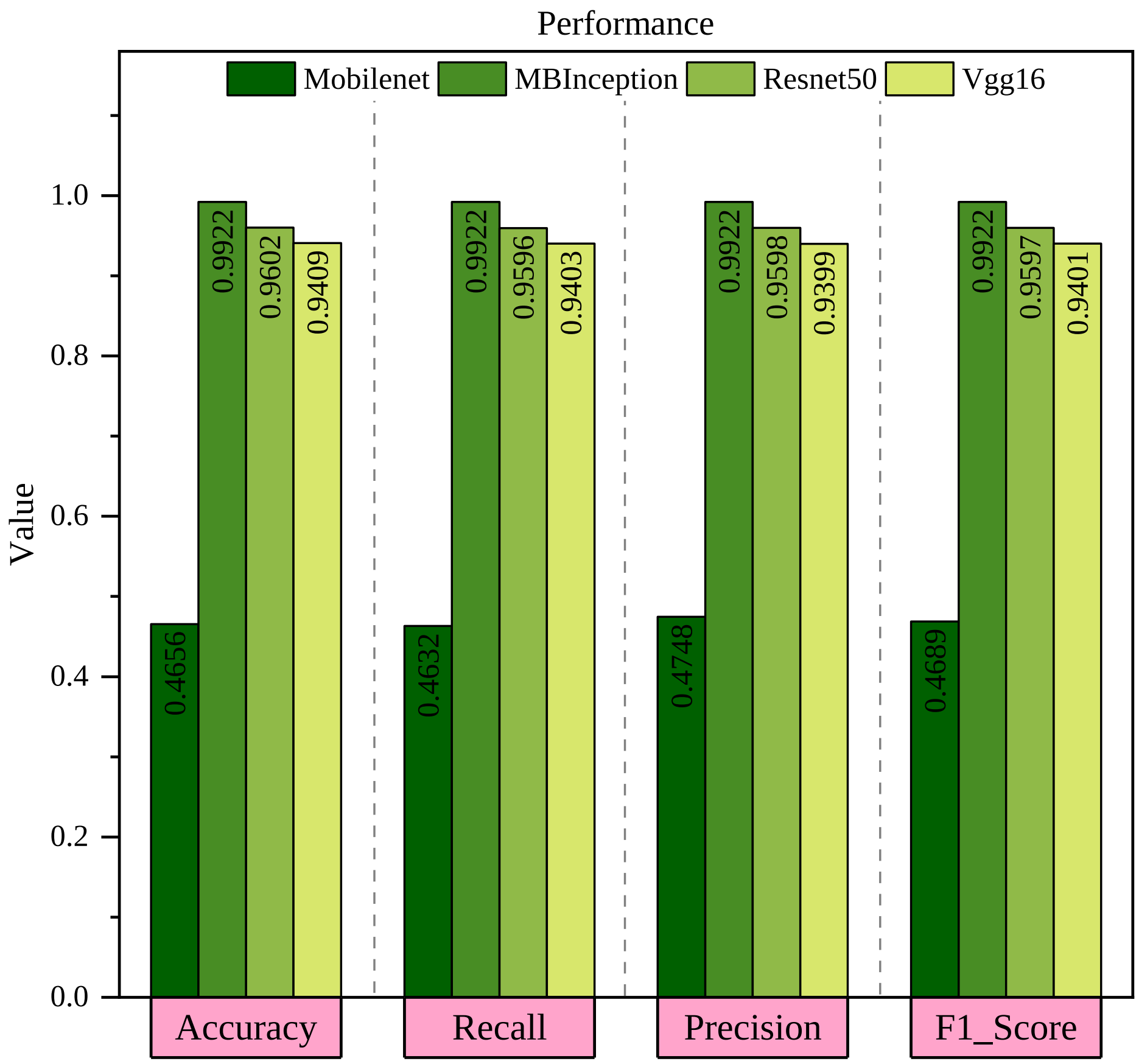

**Figure 9.** the results of evaluation metrics of models for MNIST dataset

The comparison of models based on evaluation metrics for mnist dataset (Fig. 9) show that our proposed model outperforms other models, remarkably. Metrics such as accuracy, precision, recall and f1-score denote that the MBInception has achieved high values and presented promising performance based on all the metrics.

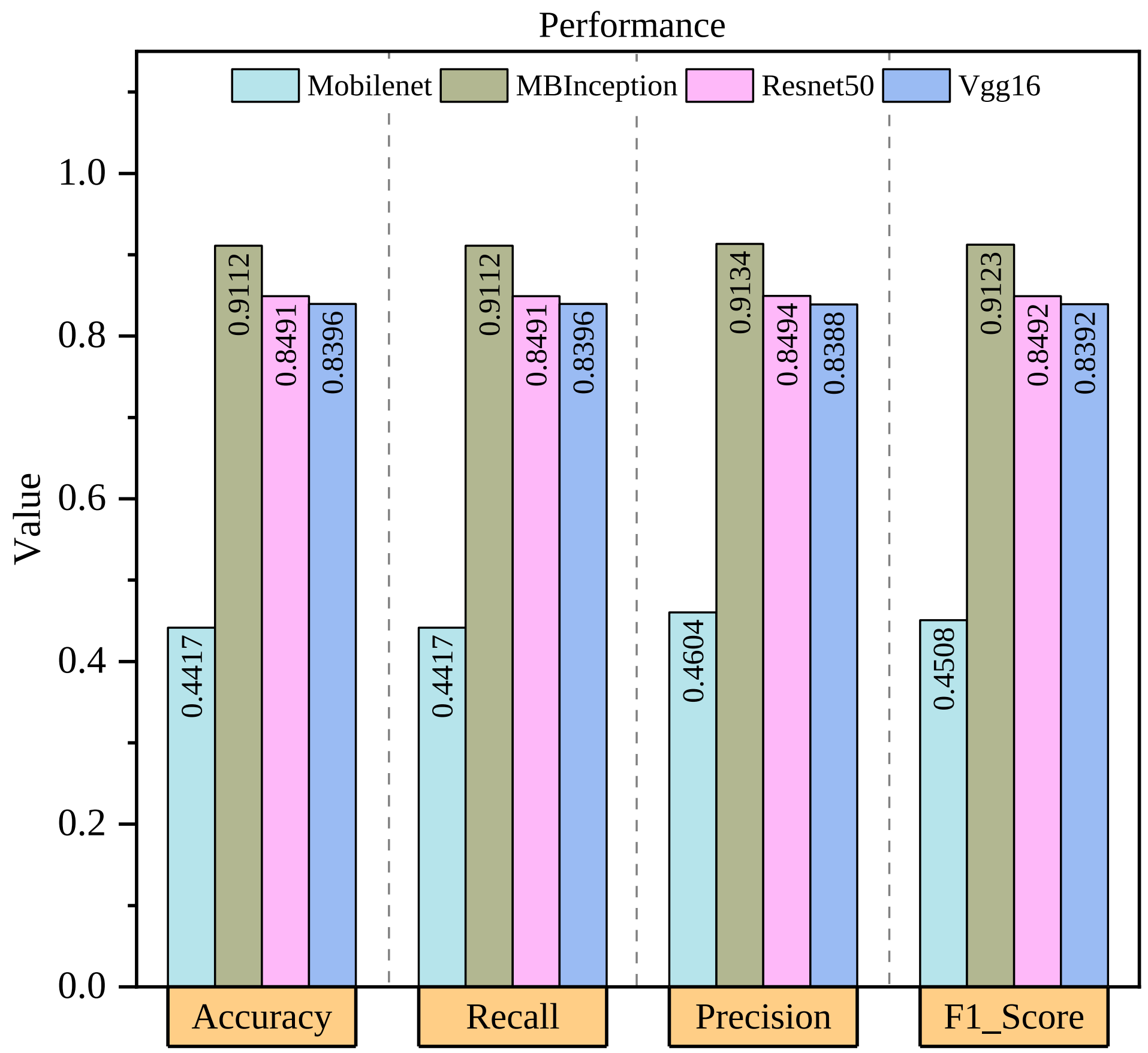

**Figure 10.** the results of evaluation metrics of models for Fashion-MNIST dataset

According to Fig. 10, By comparing the models based on evaluation criterions in fashion-mnist dataset, it is clear that the proposed model in this paper demonstrates better performance than the remaining studied models. The proposed model has improved the results of both ResNet and vgg16, explicitly. Also, the results depict that the Moobilenet shows unsatisfying results on all the datasets. In total, the proposed model has been achieved promising results in all the selected benchmark datasets and enhanced the image classification and image processing efficiency.

## 4. Conclusion

Using deep learning in image classification is a recent field of study. Nonetheless, there are numerous studies accomplished in this area. Various deep learning models are introduced with different structures and algorithms. Convolutional Neural Networks (CNNs) are a class of deep neural networks used for image recognition and image classification. Some of the common CNN-based models such as ResNet (Residual Network), VGG (Visual Geometry Group) and MobileNet are studied and compared in this paper. The aforementioned models are created based on Inception architecture introduced by researchers at Google at 2017.

Besides, this paper proposes a novel model using multiple blocks of inception, which is called MBInception (Multi-Block Inception). The novel model puts three blocks of inception back-to-back with two inception modules in each block, where tasks such as batch normalization, maxpooling, drop out and activation are implemented on unput data.

Four benchmark datasets are selected to be used for train, test and validation of models. The first one is cifar10, where vgg16 and then ResNet, showed better performance, partially. But for the other datasets including cifar100, MNIST, and fashion-MNIST, the newly proposed model demonstrated better results and overcomes models like vgg16 and ResNet. the MBInception model predicted most of the labels especially in MNIST and fashion-MNIST correctly and presented higher confidence. This is accomplished while the proposed model has less parameters compared to ResNet model.

According to evaluation criterions including accuracy, precision, recall and f1-score, the newly proposed model obtained the most satisfying results for cifar100, MNIST and fashion-MNIST datasets and succeeded over the other models. Howsoever, in cifar10 dataset, vgg16 and ResNet, were slightly better in term of evaluation criterions.

**Statements and Declarations**

☒ The authors declare that they have no known competing financial interests or personal relationships that could have appeared to influence the work reported in this paper.

**Data Availability Statements**

The datasets used in this study, including CIFAR-10, CIFAR-100, MNIST, and Fashion-MNIST, are publicly available and widely used in the fields of machine learning and computer vision.

- The **CIFAR-10** and **CIFAR-100** datasets are available through the Canadian Institute for Advanced Research (CIFAR) and can be accessed at https://www.cs.toronto.edu/~kriz/cifar.html.
- The **MNIST** dataset is available through the Modified National Institute of Standards and Technology and can be accessed at http://yann.lecun.com/exdb/mnist/.
- The **Fashion-MNIST** dataset is publicly available and can be accessed at https://github.com/zalandoresearch/fashion-mnist.

All datasets are freely accessible, and no restrictions apply to their use.

## References

Affonso, C., Rossi, A. L. D., Vieira, F. H. A., & de Carvalho, A. C. P. de L. F. (2017). Deep learning for biological image classification. *Expert Systems with Applications*, *85*, 114–122. https://doi.org/10.1016/j.eswa.2017.05.039

Albawi Saad, Abed MOHAMMED Tareq, & Al-Zawi Saad. (2017). *Understanding of a Convolutional Neural Network*.

Barz, B., & Denzler, J. (2020). Do We Train on Test Data? Purging CIFAR of Near-Duplicates. *Journal of Imaging*, *6*(6). https://doi.org/10.3390/JIMAGING6060041

Benhari, M., & Hossseini, R. (2022). An Improved Fuzzy Deep Learning (IFDL) model for managing uncertainty in classification of pap-smear cell images. *Intelligent Systems with Applications*, *16*. https://doi.org/10.1016/j.iswa.2022.200133

Dino, H., Abdulrazzaq, M. B., Zeebaree, S. R. M., Sallow, A. B., Zebari, R. R., Shukur, H. M., & Haji, L. M. (2020). *Facial Expression Recognition based on Hybrid Feature Extraction Techniques with Different Classifiers*.

Dozat, T. (2018). *INCORPORATING NESTEROV MOMENTUM INTO ADAM*.

Gholamalinezhad, H., & Khosravi, H. (n.d.). *Pooling Methods in Deep Neural Networks, a Review*.

Halgamuge, M. N., Daminda, E., & Nirmalathas, A. (2020). Best optimizer selection for predicting bushfire occurrences using deep learning. *Natural Hazards*, *103*(1), 845–860. https://doi.org/10.1007/s11069-020-04015-7

Hassan, S. U., Ahamed, J., & Ahmad, K. (2022). Analytics of machine learning-based algorithms for text classification. *Sustainable Operations and Computers*, *3*, 238–248. https://doi.org/10.1016/j.susoc.2022.03.001

He, K., Zhang, X., Ren, S., & Sun, J. (2015). *Deep Residual Learning for Image Recognition*. http://arxiv.org/abs/1512.03385

Howard, A. G., Zhu, M., Chen, B., Kalenichenko, D., Wang, W., Weyand, T., Andreetto, M., & Adam, H. (2017). *MobileNets: Efficient Convolutional Neural Networks for Mobile Vision Applications*. http://arxiv.org/abs/1704.04861

Hua, W., Zhang, Y., Liu, H., Xie, W., & Jin, X. (2024). Multichannel semi-supervised active learning for PolSAR image classification. *International Journal of Applied Earth Observation and Geoinformation*, *127*. https://doi.org/10.1016/j.jag.2024.103706

Kadam, S. S., Adamuthe, A. C., & Patil, A. B. (2020). CNN Model for Image Classification on MNIST and Fashion-MNIST Dataset. *Journal of Scientific Research*, *64*(02), 374–384. https://doi.org/10.37398/jsr.2020.640251

Krizhevsky, A. (n.d.). *Convolutional Deep Belief Networks on CIFAR-10*.

LeCun Y, Cortes C, & Burges C. (2010). MNIST Handwritten Digit Database. *AT & T Labs*, *2*.

Li Zewen, Yang Wenjie, Peng Shouheng, & Liu Fan. (2004). *A Survey of Convolutional Neural Networks*.

Mikołajczyk Agnieszka, & Grochowski Michal. (2018). *Data augmentation for improving deep learning in image classification problem*.

Okolo, G. I., Katsigiannis, S., & Ramzan, N. (2022). IEViT: An enhanced vision transformer architecture for chest X-ray image classification. *Computer Methods and Programs in Biomedicine*, *226*. https://doi.org/10.1016/j.cmpb.2022.107141

O'Shea, K., & Nash, R. (2015). *An Introduction to Convolutional Neural Networks*. http://arxiv.org/abs/1511.08458

Sharma, S., Sharma, S., & Athaiya, A. (2020). ACTIVATION FUNCTIONS IN NEURAL NETWORKS. In *International Journal of Engineering Applied Sciences and Technology* (Vol. 4). http://www.ijeast.com

Simonyan, K., & Zisserman, A. (2014). *Very Deep Convolutional Networks for Large-Scale Image Recognition*. http://arxiv.org/abs/1409.1556

Stollenga, M., Masci, J., Gomez, F., & Schmidhuber, J. (2014). *Deep Networks with Internal Selective Attention through Feedback Connections*. http://arxiv.org/abs/1407.3068

Sun, M., Song, Z., Jiang, X., Pan, J., & Pang, Y. (2017). Learning Pooling for Convolutional Neural Network. *Neurocomputing*, *224*, 96–104. https://doi.org/10.1016/j.neucom.2016.10.049

Szegedy, C., Liu, W., Jia, Y., Sermanet, P., Reed, S., Anguelov, D., Erhan, D., Vanhoucke, V., & Rabinovich, A. (2014). *Going Deeper with Convolutions*. http://arxiv.org/abs/1409.4842

Thakur, R., & Panse, P. (2022). ELSET: Design of an Ensemble Deep Learning Model for improving satellite image Classification Efficiency via Temporal Analysis. *Measurement: Sensors*, *24*. https://doi.org/10.1016/j.measen.2022.100437

Wang, F., Jiang, M., Qian, C., Yang, S., Li, C., Zhang, H., Wang, X., & Tang, X. (2017). *Residual Attention Network for Image Classification*.

Wang, P., Fan, E., & Wang, P. (2021). Comparative analysis of image classification algorithms based on traditional machine learning and deep learning. *Pattern Recognition Letters*, *141*, 61–67. https://doi.org/10.1016/j.patrec.2020.07.042

Xiao, H., Rasul, K., & Vollgraf, R. (2017). *Fashion-MNIST: a Novel Image Dataset for Benchmarking Machine Learning Algorithms*. http://arxiv.org/abs/1708.07747

Yang, X., Ye, Y., Li, X., Lau, R. Y. K., Zhang, X., & Huang, X. (2018). Hyperspectral image classification with deep learning models. *IEEE Transactions on Geoscience and Remote Sensing*, *56*(9), 5408–5423. https://doi.org/10.1109/TGRS.2018.2815613

Yang, Y., Wu, Q. M. J., Feng, X., & Akilan, T. (2020). Recomputation of the dense layers for performance improvement of DCNN. *IEEE Transactions on Pattern Analysis and Machine Intelligence*, *42*(11), 2912–2925. https://doi.org/10.1109/TPAMI.2019.2917685

Zhang, K., Sun, M., Han, T. X., Yuan, X., Guo, L., & Liu, T. (2016). *Residual Networks of Residual Networks: Multilevel Residual Networks*. https://doi.org/10.1109/TCSVT.2017.2654543